\documentclass[aps,prb,twocolumn,groupedaddress]{revtex4}
\usepackage{amsmath}
\usepackage{amssymb}
\usepackage{graphicx}

\begin{document}


\title{From spin and orbital SU(4) to spin SU(2) Kondo effect in double quantum dot}


\author{A. L. Chudnovskiy}
\email[]{achudnov@physik.uni-hamburg.de}
\affiliation{1. Institut f\"ur Theoretische Physik, Universit\"at Hamburg,
Jungiusstr. 9, 20355 Hamburg, Germany}

\author{Frank Hellmuth}
\email[]{fhellmut@physnet.uni-hamburg.de}
\affiliation{1. Institut f\"ur Theoretische Physik, Universit\"at Hamburg,
Jungiusstr. 9, 20355 Hamburg, Germany}

\author{V. Kagalovsky}
\email[]{victork@sce.ac.il}
\affiliation{Sami Shamoon College of Engineering, 
Israel}

\date{\today}

\begin{abstract}
We consider spin and orbital Kondo effect in a
parallel arrangement of two strongly electrostatically coupled quantum dots.
Increasing the exchange of electrons between the dots through the attached leads
induces a smooth crossover between $SU(4)$ spin- and orbital Kondo effect and
$SU(2)$ spin Kondo effect. Being the same for the
$SU(4)$ and $SU(2)$ symmetry points, the Kondo temperature drops slightly in the intermediate regime.
Experimentally, two kinds of Kondo effect can be discriminated by the
sensitivity to the suppression of the spin Kondo effect by Zeeman field. The
dependence of the Kondo temperature and of the differential conductance on
the strength of electronic exchange through the leads and Zeeman field is
analyzed in detail.
\end{abstract}


\maketitle


\section{Introduction}

Kondo effect in quantum dot devices remains a subject of active theoretical and experimental
investigations since its first observation \cite{DGG}. 
The observation of the Kondo effect with orbitally degenerate levels provided
the demonstration of a strong influence of the orbital structure of the
states in the dot and attached leads on the Kondo effect \cite{Kouwenhoven2}.
Further exploration of the interplay between spin and orbital degrees of freedom
in the Kondo effect became possible in experiments with double quantum dot
systems.
If the two dots are strongly electrostatically coupled \cite{Holleitner,Weis},  then
there are regions in the charging diagram of the double dot device, where there is no energy cost to transfer an electron between the two dots. In that regions, the two ground states
of the double dot system with occupations of the two dots $(N_1-1, N_2+1)$ and $(N_1, N_2)$
are degenerate. Those ground states span a two-dimensional Hilbert space
of the representation of the $SU(2)$ group, hence a spin-like degree of
freedom called pseudospin can be assigned to them. Quite analogously
to and to a large extent independently of the well-known spin Kondo effect,  the
orbital fluctuations in transport through the double quantum dot result in the
development of the orbital, or pseudospin Kondo effect.
Furthermore, at special values of Kondo couplings, the combined spin
and pseudospin Kondo Hamiltonian possesses a $SU(4)$ symmetry with respect to
rotations in spin-pseudospin space. In that regime, the $SU(4)$  Kondo effect
with greatly enhanced Kondo temperature has been predicted theoretically
\cite{Borda,Zarand}. Another realization of the $SU(4)$ Kondo effect has been
predicted theoretically for a system of a small quantum dot and a large grain. 
The signature of the $SU(4)$ Kondo effect in that system is a complete smearing
of the Coulomb blockade oscillations in the grain \cite{LeHur-Simon}. 
Recently, spin and orbital $SU(4)$ Kondo effect has been observed experimentally
in carbon nanotubes \cite{Jarillo-Herrero}. 

The existence of the pseudospin Kondo effect crucially relies on the coupling of each localized orbital state to its own reservoir of delocalized electrons. In the case of the double quantum dot, the reservoirs are  formed by the electron states in the attached leads. The separation of the reservoirs allows to define a pseudospin for the electrons in the leads in a natural way. For two sequentially coupled quantum dots \cite{Borda}, or for a double dot in the Aharonov-Bohm configuration \cite{Lopez1} such a separation is given by geometry. In contrast, the realistic experimental geometry for two dots coupled {\it in parallel} to the leads does not introduce a separate reservoir for each dot  {\it a priori} \cite{Holleitner}.

In this paper, we investigate the possibility and properties of spin and
pseudospin Kondo effect in a double dot embedded in a parallel circuit with the
attached leads if there is no separate electron reservoirs for each quantum dot.
The separation between the two electronic reservoirs is characterized by the
asymmetry of amplitudes of each electronic mode to tunnel from the reservoir
into one or into the other dot.  We introduce a measure of mixing between the
two reservoirs, which reflects the structure of tunnel matrix elements. In the
limit of completely symmetric tunneling from each mode to both quantum dots, the
Kondo effect in the orbital sector is suppressed. At the same time, all modes of
both reservoirs couple to a single electron state given by a symmetric
combination of the states localized in each dot. This results in the doubling of
the density of states in the reservoir for the spin Kondo effect. At the end,
the Kondo temperatures for the two symmetry points, the $SU(4)$ one and the
$SU(2)$ one, are equal \cite{Zarand,Lopez1,Choi,Martins}. We find however that 
the
Kondo temperature gets slightly suppressed in the intermediate regime between
the two symmetry points (see inset in Fig. \ref{figTk-b1}).  The change in
symmetry of the Kondo effect can be detected experimentally by applying an
external Zeeman magnetic field. To exclude the effect of magnetic field on the
orbital electron motion \cite{Lopez1}, the magnetic field should be applied  in
the plane of the double dot device. An external Zeeman field suppresses the
Kondo effect in the spin sector. Therefore, whereas the $SU(2)$ Kondo effect is
completely suppressed by the Zeemann field, the combined spin-pseudospin
$SU(4)$ Kondo effect is only reduced to the $SU(2)$ Kondo effect in the
pseudospin sector. 

The rest of the paper is organized as follows: the theoretical model is
introduced in Section \ref{sec-model}. In Section \ref{secT} the
renormalization group (poor man scaling) treatment is introduced.  The
results for the Kondo temperature as a function of mixing between the reservoirs
and of a Zeeman magnetic field in the whole range between the $SU(4)$ and
$SU(2)$ regimes are discussed. This part is an extension of the previous work of
one of the authors \cite{Chudnovskiy}.  The results on the conductance
of the double quantum dot are presented in Section \ref{sec-cond}. Our findings
are summarized in Section \ref{sec-conclusions}.

\section{The model of double quantum dot}
\label{sec-model}

We consider a device consisting of two single-level quantum dots coupled in parallel to
external Fermi liquid leads. The Hamiltonian represents a sum of the following
terms: the Hamiltonian of isolated quantum dots including the interdot
interaction, the Hamiltonian of Fermi liquid reservoirs,
and the tunneling between the reservoirs and the dots
\begin{equation}
H=H_{QD}+\sum_{\nu=r,l}H^{\rm res}_{\nu}+H^t.
\label{H1}
\end{equation}
The Hamiltonian of isolated quantum dots reads
\begin{equation}
H_{QD}=\sum_{i=1,2}\left[\sum_{\sigma=\uparrow\downarrow}E_i\hat{n}_{i\sigma}
+U\hat{n}_{i\uparrow}\hat{n}_{i\downarrow}\right]+U_{12}\hat{n}_1\hat{n}_2,
\label{Hqd}
\end{equation}
where the following notations are introduced: $\hat{n}_{i\sigma}=
\hat{c}^{\dagger}_{i\sigma}\hat{c}_{i\sigma}$ -- the number of electrons in the dot
$i$ with spin-projection $\sigma$, $\hat{c}_{i\sigma}$ is an annihilation operator of
an electron in dot $i$ with spin projection $\sigma$. We focus on the $SU(4)$ Kondo
regime that  is achieved at $E_1=E_2=E$, $U_{12}=U$.
The low energy sector of the model consists of states with a total of one electron in the
double dot. Their energy in the isolated dots equals $E \ (<0)$.
The corresponding ket-vectors are denoted as $|\sigma,0\rangle$ and
$|0,\sigma\rangle$, where $\sigma=\uparrow \mbox{or} \downarrow$ refers to the
$z$-projection of spin, and $0$ denotes the unoccupied dot. Single particle
tunneling transforms the low energy states into the empty state $|0,0\rangle$
(energy $0$) or into states with a total of two electrons (energy $E+U \ (>0)$),
which are $|\uparrow\downarrow,0\rangle$,
$|0,\uparrow\downarrow\rangle$, $|\sigma_1,\sigma_2\rangle$.
The excited states with a total of more than two electrons in the double dot do not couple
to the low-energy sector in the second order in tunneling. They are omitted from
consideration.
The tunneling Hamiltonian is given by
\begin{equation}
H^t=\sum_{i=1,2}\sum_{r=s,d}\sum_{k\sigma}
t_{ik}^{r}\hat{a}^{\dagger}_{r k\sigma}\hat{c}_{i\sigma}+h.c., 
\label{tunneling}
\end{equation}
where $\hat{a}_{r k\sigma}$ is the annihilation operator of a fermion in the reservoir $r$ with the wave vector $k$ and $z$-projection of spin $\sigma=\uparrow,\downarrow$. 
We do not consider the effect of a perpendicular magnetic field throughout this
work, then the tunnel matrix elements can be chosen real. To elucidate the
appearance of the pseudospin in a general case, when each electronic mode has 
nonvanishing tunneling amplitudes to both quantum dots, let us define the
tunneling angle $\eta_k^{r}$ for each mode in the given reservoir by the
following relations:
\begin{eqnarray}
&& 
\cos\left(\eta_k^{r}\right)=\frac{t^{r}_{1k}}{t^{r}_k},
\label{coseta} \\
&&
\sin\left(\eta_k^r\right)=\frac{t^{r}_{2k}}{t^{r}_k},
\label{sineta}
\end{eqnarray}
where $t^{r}_k=\sqrt{\left(t^{r}_{1k}\right)^2
+\left(t^{r}_{2k}\right)^2}$. In what follows we consider the case of symmetric
couplings to both reservoirs, $\eta^s_k=\eta^d_k$ and omit the index of the 
reservoir by the tunneling angle. 
Furthermore, we introduce formally bi-spinor notations for the operators in the fermionic reservoir and
in the double dot
\begin{eqnarray}
&&
\hat{\Psi}_{r k}=\left(\cos\eta^{r}_k, 
\sin\eta^{r}_k\right)^T\otimes
\left(\hat{a}_{r k\uparrow},\hat{a}_{r k\downarrow}\right)^T , \label{Psi}\\
&&
\hat{\Phi}=\left(\hat{c}_{1\uparrow},\hat{c}_{1\downarrow},
\hat{c}_{2\uparrow},\hat{c}_{2\downarrow}\right)^T. \label{Phi}
\end{eqnarray}
In (\ref{Psi}) the upper signs in the exponent relate to $r=s$ and the lower ones to $r=d$.
Using the notations above,  we rewrite the tunneling Hamiltonian in the form
\begin{equation}
 H^t=\sum_{k}t_k\left(\frac{t_k^s}{t_k}\hat{\Psi}_{sk}^{\dagger}+
 \frac{t_k^d}{t_k}\hat{\Psi}_{dk}^{\dagger}\right)\hat{\Phi}+ h.c.,
\label{HT}
\end{equation}
where $t_k=\sqrt{(t_k^s)^2+(t_k^d)^2}$.
One can see from (\ref{HT}) that for each $k$ it is only the mode 
\begin{equation}
\hat{\Psi}_k=\frac{t_k^s}{t_k}\hat{\Psi}_{sk}
+ \frac{t_k^d}{t_k}\hat{\Psi}_{dk}
\label{Psi_k}
\end{equation}
that couples to the double dot \cite{Glazman-Raikh}.
Performing the Schriefer-Wolff transformation for Hamiltonian (\ref{H1}), we derive the effective Kondo Hamiltonian describing the dynamics of the low-energy sector of the model that can be written as
\begin{equation}
H_K=\sum_{\mu,\nu=0}^3\sum_{k,k'} J^{kk'}_{\mu\nu}\hat{\Phi}^{\dagger}\left(\hat{\tau}^{\mu}\otimes\hat{\tau}^{\nu}\right)
\hat{\Phi}\hat{\Psi}_{k}^{\dagger}\left(\hat{\sigma}^{\mu}\otimes\hat{\sigma}^{\nu}\right)
\hat{\Psi}_{k'}+H_{\mathrm{res}}.
\label{HK}
\end{equation}
Here $\hat{\Phi}^{\dagger}\left(\hat{\tau}^{\mu}\otimes\hat{\tau}^{\nu}\right)
\hat{\Phi}$ and $\hat{\Psi}_k^{\dagger}\left(\hat{\sigma}^{\mu}\otimes\hat{\sigma}^{\nu}\right)
\hat{\Psi}_k$ represent the ``hyperspins'' in the pseudospin-spin $SU(4)$ space
for the double dot and fermionic reservoir respectively, $\hat{\tau}^{\mu}$ and
$\hat{\sigma}^{\mu}$
denote the corresponding Pauli matrices. For the $SU(4)$ symmetry point, the
Kondo couplings equal
$J_{\mu\nu}^{kk'}= t_kt_{k'}\left(\frac{1}{E+U}-\frac{1}{E}\right)$ for all
$\mu,\nu=\overline{0,3}$ except
$\mu=\nu=0$.
Note the principal difference between the hyperspins in the quantum
dot and in the reservoir. Whereas the operator $\hat{\Phi}$ incorporates four
dynamical degrees of freedom,
the operator $\hat{\Psi}_k$ does only two, $\hat{a}_{k\uparrow}$ and $\hat{a}_{k\downarrow}$. The angle $\eta_k$ is not dynamical. Yet this angle can fluctuate with wave vector $k$.
The magnitude of those fluctuations determines whether the pseudospin in reservoirs is promoted to a
dynamical degree of freedom. For example, if the reservoirs for two quantum dots are strictly separated, the angle $\eta_k$ assumes two possible values $0$ and $\pi/2$, which results in strong fluctuations
of $\eta_k$ with $k$, and eventually leads to the pseudospin Kondo effect. In contrast, if there is only a single common reservoir for both dots, then the tunneling amplitudes to both dots are equal
$t_{1k}=t_{2k} \ \forall k$. In that case, the angle $\eta_k$ is frozen at the
value $\pi/4$ for any $k$,
and the orbital Kondo effect is suppressed.

\section{Kondo temperature \label{secT}}

\begin{figure}
\includegraphics[width=4cm,height=1.5cm,angle=0]{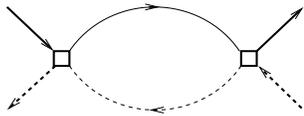}
\caption{One loop renormalization of Kondo coupling.  Solid line: the propagator of
electrons in the reservoir $\hat{G}(\omega,k)$. Dashed line: the spin-propagator in the double
dot $\hat{D}(\omega)$, see (\ref{D}), (\ref{G}).
 \label{fig-1loop}}
\end{figure}

In what follows we present results on the dependence of Kondo temperature on
the mixing between the reservoirs $b_1$ obtained by  the renormalization
group (or poor man scaling) treatment of  the Hamiltonian (\ref{HK}). We assume
the values $t^r_k$
that reflect  the total tunneling probability from a given mode of a reservoir
into the double dot to be $k$-independent for energies close to the Fermi level
of the reservoirs, $t^r_k=t^r$. Then the Kondo coupling constants
$J^{kk'}_{\mu\nu}$ become independent of the wave vectors, 
$J^{kk'}_{\mu\nu}=J_{\mu\nu}$.  The one loop renormalization of the Kondo
couplings $J_{\mu\nu}$ is given by the diagram in Fig. \ref{fig-1loop} and the
diagrams obtained by change of direction of arrows for solid or dashed lines.
The propagators of the field in the double dot $\hat{D}(i\omega_n)$ and
in the electronic reservoir $\hat{G}(i\omega_n,k)$ are given by
\begin{eqnarray}
&&
\hat{D}(i\omega_n)=\frac{1}{\left(i\omega_n\right)} 
\left(\hat{1}_2\otimes\hat{1}_2\right), \label{D} \\
\nonumber &&
\hat{G}(i\omega_n,k)=\frac{1}{i\omega_n-\xi_k}\times \\ 
&& \left[
\begin{array}{cc}
\cos^2\eta_k & \cos\eta_k\sin\eta_k \\
\cos\eta_k\sin\eta_k & \sin^2\eta_k
\end{array}
\right]\otimes \hat{1}_2. \label{G}
\end{eqnarray}
One can see that in comparison to the pure $SU(2)\otimes SU(2)$ Kondo effect as predicted for sequentially coupled dots, the presence of mixing between the reservoirs results in a nontrivial structure of the fermion propagator $\hat{G}(i\omega_n,k)$ in orbital space.
After integration in the infinitesimal energy shell between $\Lambda-\delta\Lambda$ and $\Lambda$, $\Lambda$ being the  high energy cutoff,  the correction to the Hamiltonian can be written as
\begin{widetext}
\begin{eqnarray}
\nonumber &&
\delta \hat{H}_K
=\pi\nu_F \left(\delta\Lambda/\Lambda\right) J_{\mu\nu}J_{\mu'\nu'}\hat{\Phi}^{\dagger}
\left[\left(\hat{\tau}^{\mu}
\otimes\hat{\tau}^{\nu}\right)\left(\hat{\tau}^{\mu'}\otimes\hat{\tau}^{\nu'}\right)-
\left(\hat{\tau}^{\mu'}\otimes\hat{\tau}^{\nu'}\right)\left(\hat{\tau}^{\mu}
\otimes\hat{\tau}^{\nu}\right)\right]\hat{\Phi}\times \\
&&
\hat{\Psi}^{\dagger}\left[\left(\hat{\sigma}^{\mu}\otimes\hat{\sigma}^{\nu}\right)\hat{P}
\left(\hat{\sigma}^{\mu'}\otimes\hat{\sigma}^{\nu'}\right)-
\left(\hat{\sigma}^{\mu'}\otimes\hat{\sigma}^{\nu'}\right)\hat{P}
\left(\hat{\sigma}^{\mu}\otimes\hat{\sigma}^{\nu}\right)
\right]\hat{\Psi}.
\label{deltaHK}
\end{eqnarray}
\end{widetext}
Here $\nu_F$ denotes the density of states at the Fermi level,
$
\hat{P}=\left(b_0\hat{1}_2+\sum_{p=1}^3b_p\hat{\sigma}_p\right)\otimes \hat{1}_2
\label{P}
$, $b_0=1/2$, and the values of other constants $b_p$ are given by the following averages over the wave vector  $k$
\begin{eqnarray}
&&
b_1=\langle\cos\eta_k\sin\eta_k\rangle, \label{b1}\\
&&
b_2=0, \ \mbox{for $\eta_k^s=\eta_k^d$}, \label{b2} \\
&&
b_3=\frac{1}{2}\langle\cos^2\eta_k-\sin^2\eta_k\rangle. \label{b3}
\end{eqnarray}
Due to the nontrivial structure of the fermion propagator new interactions are
generated in course of the renormalization group (RG) transformation. The new
interactions have the structure
$J_{\mu\lambda\nu}\hat{\Phi}^{\dagger}\left(\hat{\tau}^{\mu}\otimes
\hat{\tau}^{\nu} \right) \hat{\Phi} \cdot
\hat{\Psi}^{\dagger}\left(\hat{\sigma}^{\lambda}\otimes\hat{\sigma}^{\nu}\right)
\hat{\Psi}$, where all the indices $\mu$, $\lambda$
and $\nu$ change independently.

We consider the case where the tunnel couplings to the two dots are symmetric,
that is $b_3=0$. The mixing between the two reservoirs is now determined by the
value of the parameter $b_1$. The case $b_1=0$ corresponds to the case of two
strictly separated reservoirs, where the $SU(4)$ Kondo effect is expected,
whereas the maximal value $b_1=b_0=1/2$ corresponds to a single common reservoir
for both quantum dots. In the latter case only the spin $SU(2)$ Kondo effect is
possible. The chosen form of the operator $\hat{P}$ with $b_1\neq 0$
distinguishes the coupling constants with $\mu,\lambda=\overline{0,1}$. It turns
out that the only coupling constants with $\mu\neq\lambda$ that are generated by
the RG-transformation are $J_{10\nu}$ and $J_{01\nu}$. The constants with
$\mu=\lambda=\overline{2,3}$ have the same RG-flow, we further denote them as
$J_{\perp\nu}$.
Without loss of generality  we take the direction of the Zeeman magnetic field
along the $z$-axes. We use the approximation that the Zeeman field freezes out
the spin fluctuations as soon as the running value of Zeeman energy equals
the high energy cutoff $\Lambda$. The condition for the running Zeeman energy 
$h(l_h)=h_0e^{l_h}=\Lambda$ determines the logarithmic energy scale $l_h$. For
the logarithmic energy scale $l<l_h$ the influence of the Zeeman field is
neglected, all the spin components being equivalent.  For $l>l_h$ the structure
of the RG-equation changes abruptly, the couplings with the spin-index $\nu=1,2$
stop to flow.

\begin{figure}
\includegraphics[width=8cm,height=6cm,angle=0]{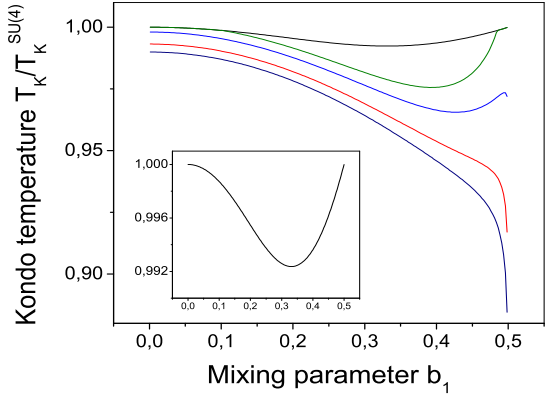}
\caption{(Color online) Kondo temperature as a function of mixing $b_1$ for
different Zeeman fields $h$. Magnetic field increases from the highest to the
lowest curve. From the highest to the lowest curve: $h/T_K^{SU(4)}=1.0; 1.02;
1.06; 1.14; 1.29$. Inset: the dependence $T_k(b_1)$ at zero Zeeman field. The
curves illustrate characteristic $T_K(b_1)$ dependencies.
\label{figTk-b1}}
\end{figure}

In what follows we absorb the density of states $\nu_F$ into the definition of coupling constants $\pi\nu_F J_{\mu\lambda\nu}\mapsto J_{\mu\lambda\nu}$. For $l<l_h$, the RG-equations for the coupling constants can be written as
\begin{eqnarray}
&&
\frac{d}{dl}Q=Q^2+R^2+\bar{J}^2, \label{eqQ} \\
&&
\frac{d}{dl}R=2QR, \label{eqR} \\
&&
\frac{d}{dl}\bar{J}=\frac{3}{2}Q\bar{J}+K\bar{J}, \label{eqbarJ} \\
&&
\frac{d}{dl}K=\bar{J}^2. \label{eqK}
\end{eqnarray}
Here the following combinations of coupling constants are introduced: 
\begin{eqnarray} 
&& 
Q=2(b_0J_{11\nu}+b_1J_{10\nu}), \label{Q} \\
&&
R=2(b_1J_{11\nu}+b_0J_{10\nu}), \label{R}  \\
&& 
\bar{J}=2\sqrt{b_0^2-b_1^2}J_{\perp\nu}, \label{barJ} \\
&&
K=b_0J_{110}+b_1J_{100}  \label{K}
\end{eqnarray}
with $\nu=\overline{0,3}$.
Numerical solution of the RG-equations (\ref{eqQ})--(\ref{eqK}) allows to
determine
approximately the Kondo temperature as a function of $b_1$. The solution is
shown in the inset in Fig. \ref{figTk-b1}.
One can see that without the external Zeeman field, or for Zeeman energy less than the Kondo temperature
$T_K^{SU(4)}$, the Kondo temperature diminishes only slightly between the SU(4)-symmetric  point $b_1=0$ and the SU(2) symmetric point $b_1=0.5$.   Therefore, the transition between the $SU(4)$ and $SU(2)$ Kondo effect can hardly be seen in the change of the Kondo temperature without the external Zeeman field \cite{Choi}.

For $l>l_h$ the RG equations have a form
\begin{eqnarray}
&&
\frac{d}{dl}{\mathcal K}={\mathcal J}^2, \label{Kh} \\
&&
\frac{d}{dl}{\mathcal J}={\mathcal K}{\mathcal J} \label{Jh}
\end{eqnarray}
with  
\begin{eqnarray}
&&
{\mathcal J}=\sqrt{b_0^2-b_1^2}(J_{\mu\mu 0}+J_{\mu\mu 3}), \ 
\mu=\overline{2,3}, \label{effJ}  \\
&&
{\mathcal K}=b_0(J_{110}+J_{113})+b_1(J_{100}+J_{103}). \label{effK}
\end{eqnarray}
Eqs. (\ref{Kh}), (\ref{Jh}) are to be solved with the initial conditions given
by the solution of (\ref{eqQ}) -- (\ref{eqK}) at $l=l_h$:
${\mathcal J}(0)=\bar{J}(l_h)$, ${\mathcal K}(0)=K(l_h)+Q(l_h)/2$.
Eqs. (\ref{Kh}), (\ref{Jh}) also describe the suppression of the Kondo
temperature in a quantum dot coupled to ferromagnetic leads \cite{Martinek}.
Analytical solution of (\ref{Kh}), (\ref{Jh}) results in the following
dependence of the Kondo temperature on the Zeeman field
\begin{equation}
T_K(h)= h\left[\frac{{\mathcal J(0)}}{{\mathcal K}(0)+\sqrt{{\mathcal K}^2(0)-{\mathcal J}^2(0)}}
\right]^{\frac{1}{\sqrt{{\mathcal K}^2(0)-{\mathcal J}^2(0)}}}.
\label{T(h)}
\end{equation}
The dependence of the Kondo temperature on the parameter $b_1$ at different
values of Zeeman magnetic field $h$ is shown in Fig. \ref{figTk-b1}.  Rising
the Zeeman field, the dependence $T_K(b_1)$ changes from the nonmonotonous one
to the monotonously falling. Interesting enough is the nonmonotonus dependence
of Kondo temperature on the mixing parameter $b_1$ at relatively small Zeeman
fields. The dip in the $T_K(b_1)$ dependence at intermediate values of $b_1$
can be explained in term of channel blocking, a common phenomenon in transport
through quantum dots \cite{channel-block}. Namely, the two projections of the
pseudospin correspond to the two transport channels through the double quantum
dot. The wave functions in the channels are given by 
symmetric and antisymmetric combinations of the wave functions of each dot. For
the sake of brevity we further denote those channels the
symmetric one and the antisymmetric one respectively. Increasing $b_1$, one
channel (the
antisymmetric one) becomes progressively decoupled from the leads. Close to the
maximal possible value $b_1=0.5$ that channel is almost decoupled, the Kondo
correlations occur only in the strongly coupled symmetric channel leading to
the spin Kondo effect. Yet there is a small probability that the antisymmetric
channel is occupied by an electron. Such an event has a huge destructive
influence on the Kondo effect. Not only is the spin Kondo effect suppressed in
the weakly coupled channel  because of a very low Kondo temperature, the Kondo
effect in the strongly coupled channel is also blocked because the Coulomb
blockade prevents the occupation of the symmetric state as long as the
antisymmetric  one is occupied. Weak coupling of the antisymmetric channel to
the leads results in
the long life time of an electron in that state, although the probability to
occupy it by tunneling from a lead is small. The competition of a long
life-time and a low occupation probability of the blocking channel leads finally
to a nonmonotonous dependence of the Kondo temperature on the mixing parameter
$b_1$. 

\begin{figure}
\includegraphics[width=8cm,height=6cm,angle=0]{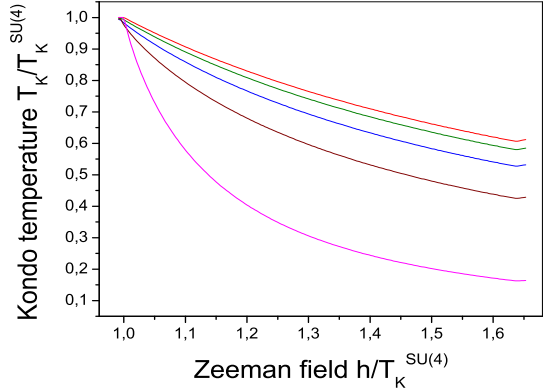}
\caption{(Color online) Kondo temperature versus Zeeman magnetic field at
different values of mixing $b_1$. 
From the highest to the lowest curve $b_1=0.1; 0.2; 0.3; 0.4; 0.49$.   \label{figTk-h}}
\end{figure}

The dependence of the Kondo temperature on Zeeman magnetic field $h$ for
different values of the mixing parameter $b_1$ is shown in Fig. 
\ref{figTk-h}. The suppression of the Kondo temperature with Zeeman field and
with the mixing $b_1$ illustrates our findings (\ref{T(h)}).

\section{Conductance of the double quantum dot device \label{sec-cond}}

\begin{figure}
\includegraphics[width=6cm,height=5cm,angle=0]{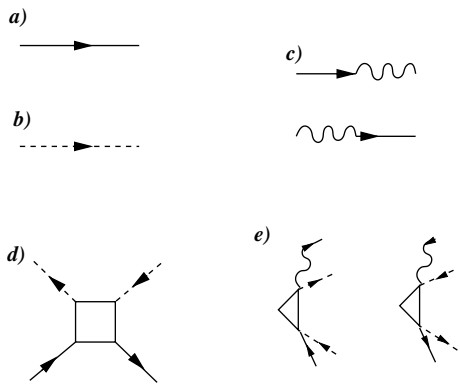}
\caption{Elements of the diagrams: a) $\mathcal{G}_0$; b) semifermion Green's
function;  
c) $\mathcal{F}_0$; d) Kondo interaction vertex; e) vertices for the current
operator.   
 \label{fig-diag-elements}}
\end{figure}

\begin{figure}
\includegraphics[width=4cm,height=3cm,angle=0]{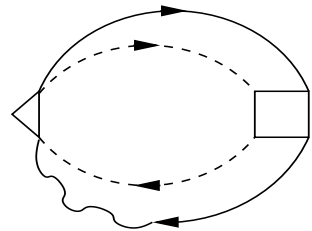}
\caption{Diagram for the lowest order nonzero contribution to the current
through the double quantum dot.   
 \label{current}}
\end{figure}

To calculate the current through the double-dot device in presence of the
applied voltage, we start with the generalization of the Kondo Hamiltonian
(\ref{HK}) taking into account the couplings that are generated in course of 
RG-transformation. The current operator is defined as an electric charge flow
out of the source lead 
\begin{eqnarray}
\nonumber &&
\hat{j}=-e\frac{\partial}{\partial t} \left(\sum_{
k}\hat{\Psi}_{k}^{s\dagger} 
\hat{\Psi}^s_{k}\right) \\
\nonumber && 
=-ie \sum_{\lambda, \mu,\nu=0}^3\sum_{p, p'}
J^{pp'}_{\lambda\mu\nu}
\hat{\Phi}^{\dagger}\left(\hat{\tau}^{\mu}\otimes\hat{\tau}^{\nu}\right)
\hat{\Phi} \times \\
&& 
 \left[\hat{\Psi}_{p}^{\dagger}\left(\hat{\sigma}^{\lambda}\otimes\hat{\sigma}
^{\nu}\right)\hat{\Psi}_{p'}, 
\sum_{k}\hat{\Psi}^{s\dagger}_{k} 
\hat{\Psi}^s_{k} \right]_{-}. 
\label{def_current}
\end{eqnarray}
Using the relation (\ref{Psi_k}) between the operators $\hat{\Psi}_{k}$ and
$\hat{\Psi}^s_{k}$, 
and evaluating the commutator in (\ref{def_current}), we can represent the current operator in the form 
\begin{equation}
\hat{j}=\hat{j}_{\mathrm{sd}}-\hat{j}_{\mathrm{ds}},
\label{def_j2}
\end{equation}
where 
\begin{eqnarray}
\nonumber &&
\hat{j}_{\mathrm{sd}}=-i e \sum_{\lambda\mu\nu} 
\sum_{k k'} J_{\lambda\mu\nu}^{k k'} 
\frac{t_{k}^d t_{k'}^s}{t_{k} 
t_{k'}} \hat{\Psi}_{k}^{d\dagger}
\left(\hat{\sigma}^{\lambda}\otimes\hat{\sigma}^{\nu}\right) 
\hat{\Psi}^s_{k'}
 \times \\ 
&&
\hat{\Phi}^{\dagger}\left(\hat{\tau}^{\mu}\otimes\hat{\tau}^{\nu}\right)
\hat{\Phi}. 
\label{j_sd}
\end{eqnarray}
We are going to calculate the current to the lowest nonzero order in the
renormalized Kondo coupling constants. To proceed with further calculation we
note that it is only the field $\hat{\Psi}_{k}$ defined by (\ref{Psi_k}) that
acquires the renormalization by Kondo interactions. Therefore, it is convenient
to express the fields $\hat{\Psi}^{s,d}_{k}$ through the field
$\hat{\Psi}_{k}$ and its orthogonal 
\begin{equation}
\hat{\Xi}_{\bf k}=\frac{t_{k}^d}{t_{k}}\hat{\Psi}^s_{k}
- \frac{t_k^s}{t_k}\hat{\Psi}^d_{k}. 
\label{Xi_k}
\end{equation}
In terms of the fields $\hat{\Psi}_{k}$ and $\hat{\Xi}_{k}$ the total
current is given by 
\begin{eqnarray}
\nonumber && 
\hat{j}=\hat{j}_{\mathrm{sd}}-\hat{j}_{\mathrm{ds}} =-i e \sum_{\lambda\mu\nu} J_{\lambda\mu\nu} \hat{\Phi}^{\dagger}\left(\hat{\tau}^{\mu}\otimes\hat{\tau}^{\nu}\right)
\hat{\Phi} \times  \\ 
\nonumber &&
\sum_{k k'} 
\frac{t_{k}^{\mathrm s} t_{k'}^{\mathrm d}}{t_{k} t_{k'}} 
\left[\hat{\Psi}_{k}^{\dagger}
\left(\hat{\sigma}^{\lambda}\otimes\hat{\sigma}^{\nu}\right) \hat{\Xi}_{k'}-
\hat{\Xi}^{\dagger}_{k}
\left(\hat{\sigma}^{\lambda}\otimes\hat{\sigma}^{\nu}\right) \hat{\Psi}_{
k'}\right].\\
\label{j_total} 
\end{eqnarray}
The current at finite transport voltage is calculated using Keldysh formalism.  
The unperturbed Green's functions  in the source and in the drain read 
\begin{equation}
\hat{G}_{\mathrm{s,d}}(\epsilon, k)=\left(
\begin{array}{cc}
G^{R}_{\mathrm{s,d}}(\epsilon,  k) & G^{K}_{\mathrm{s,d}}(\epsilon, k)
\\
0 & G^{A}_{\mathrm{s,d}}(\epsilon, k)
\end{array}
\right),
\label{Gsd}
\end{equation}
where 
\begin{eqnarray} 
&& 
G^{R/A}_{\mathrm{s,d}}(\epsilon,
k)=\frac{1}{\epsilon-\xi^{\mathrm{s,d}}_{k} \pm i
o}\hat{\Theta}_{\mathrm{s,d}}(k)   \label{Gsd_R/A} \\
\nonumber &&
G^{K}_{\mathrm{s,d}}(\epsilon, k)=\tanh\left(\frac{\epsilon-
\mu_{\mathrm{s,d}}}{2T}\right) \left(G^{R}_{\mathrm{s,d}}(\epsilon, k)
-G^{A}_{\mathrm{s,d}}(\epsilon, k)\right), \\ \label{Gsd_K}
\end{eqnarray}
where $\hat{\Theta}_{\mathrm{s,d}}(k)$ is the matrix characterizing the
tunneling angles for each mode 
\begin{equation}
\hat{\Theta}_{\mathrm{s,d}}(k)=\left(
\begin{array}{cc}
\cos^2\eta^{\mathrm{s,d}}_{k} &
\cos\eta^{\mathrm{s,d}}_{k} \sin\eta^{\mathrm{s,d}}_{k} \\
\cos\eta^{\mathrm{s,d}}_{k}\sin\eta^{\mathrm{s,d}}_{k} &
\sin^2\eta^{\mathrm{s,d}}_{k}
\end{array}
\right)\otimes \hat{1}_2.
\end{equation}
Using  relations
(\ref{Psi_k}), (\ref{Xi_k}), we obtain the following set of unperturbed Green's
functions 
\begin{eqnarray}
&&
\mathcal{G}_0=-i\left\langle\hat{\Psi}\hat{\Psi}^{\dagger}\right\rangle_0=
\frac{(t^s)^2}{t^2} G_s +\frac{(t^d)^2}{t^2}G_d, \label{PsiPsi} \\ 
\nonumber &&
\mathcal{F}_0=-i\left\langle\hat{\Xi}\hat{\Psi}^{\dagger}\right\rangle_0=-i\left
\langle \hat{\Psi}
\hat{\Xi}^{\dagger}\right\rangle_0 \\ 
&& =\frac{t^s t^d}{t^2}\left(G_s -G_d\right),
\label{PsiXi}\\
&&
g_0=-i\left\langle\hat{\Xi}\hat{\Xi}^{\dagger}\right\rangle_0=\frac{(t^d)^2}{t^
2}G_s +\frac{(t^s)^2}{t^2}G_d. \label{XiXi}  
\end{eqnarray}
Here we suppressed the spatial and temporal arguments of the Green's  functions
as well as their indices in Keldysh space. 
Now we calculate the Kondo correction to the current through the double dot
using the  diagrammatic expansion based on the Green's functions (\ref{PsiPsi})
-- (\ref{XiXi}). Note that according to the structure of the Kondo interaction
term in (\ref{HK}) it is only the Green function (\ref{PsiPsi}) that acquires
corrections from the Kondo interaction. The diagram elements are represented in
Fig. \ref{fig-diag-elements}. The lowest order nonzero correction to the current
is given by the diagram in Fig. \ref{current}. 

\begin{figure}
\includegraphics[width=3.5cm,height=2.cm,angle=0]{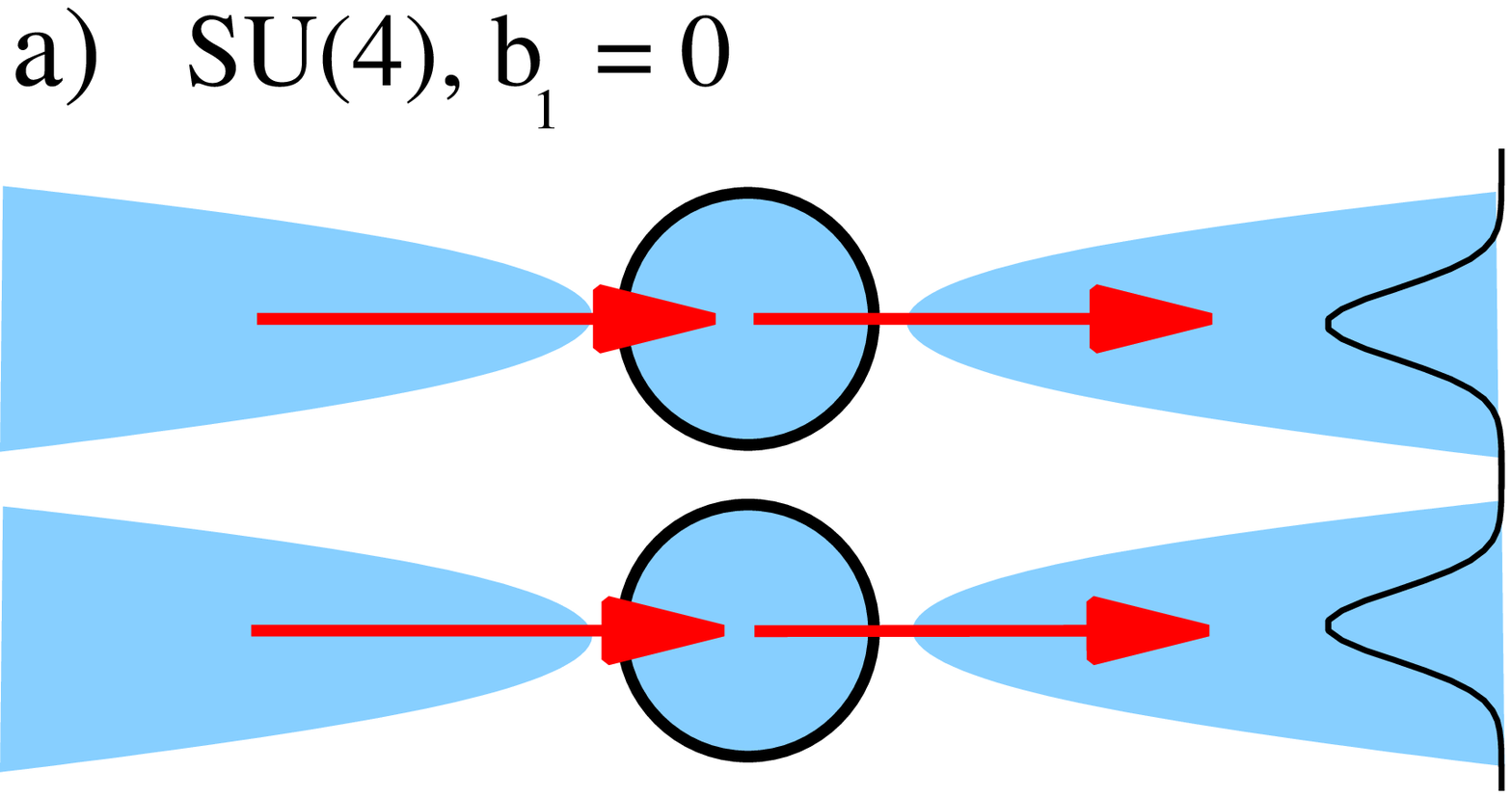} \hskip 0.5cm 
\includegraphics[width=3.5cm,height=2.cm,angle=0]{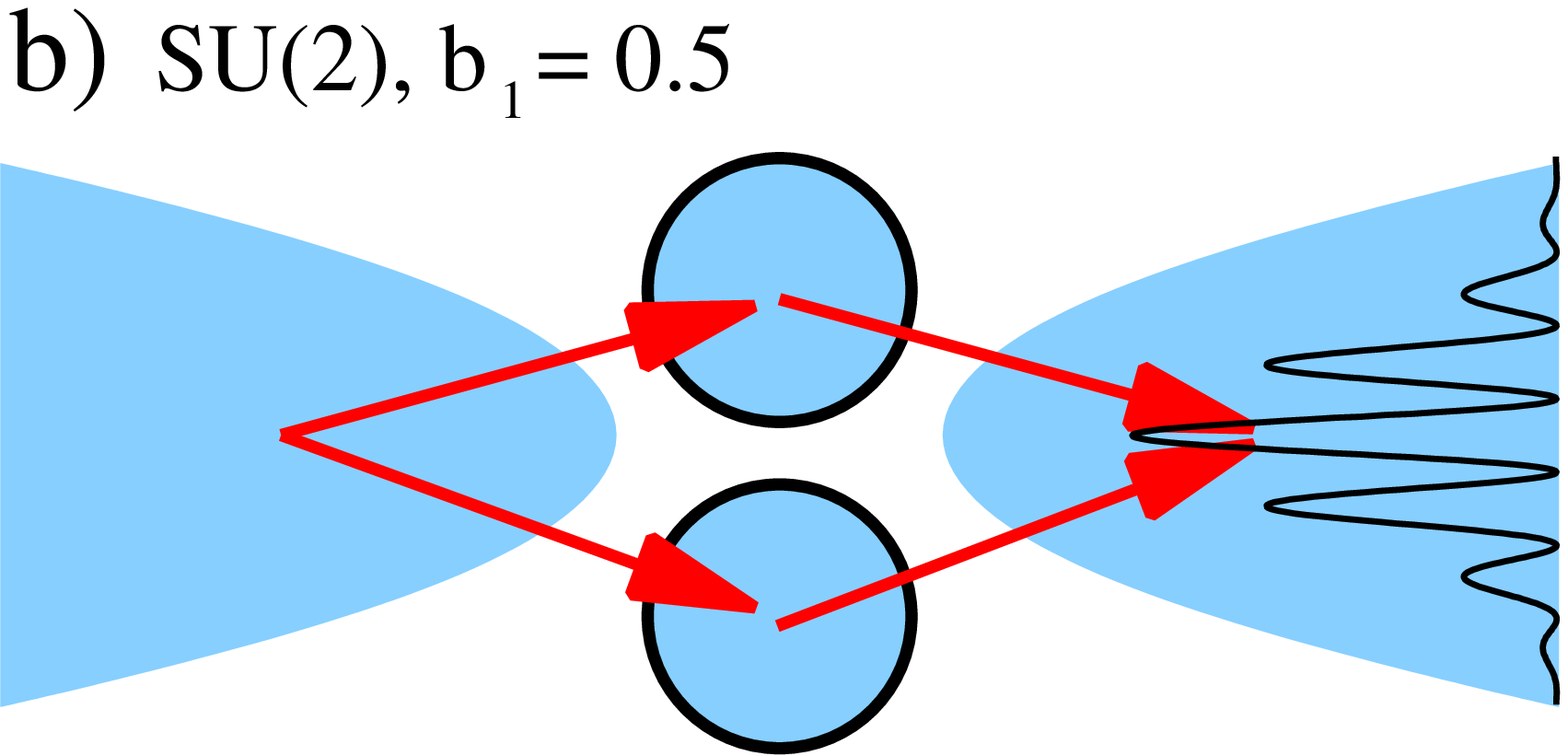}
\caption{(Color online) a) Scheme of conducting channels in the $SU(4)$ Kondo 
effect. There are two conductance channels with correlated transport but
without interference. The total current is given by the sum of currents through
each channel. b) Scheme of conducting channels in the spin $SU(2)$ Kondo effect
in double quantum dot. The double slit geometry is realized. There is a single 
channel that is splitted between the two quantum dots. The total current is
given by interference of the two partial waves in the central maximum.  
 \label{fig-dd-channels}}
\end{figure}

\begin{figure}
\includegraphics[width=6cm,height=5cm,angle=0]{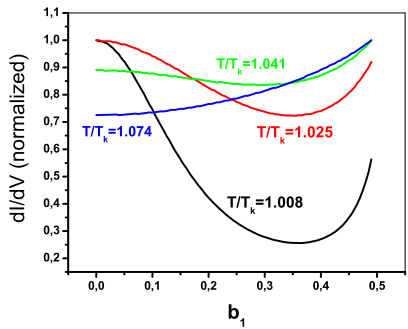}
\caption{(Color online) Differential conductance of the double quantum dot vs
the mixing parameter $b_1$ at different temperatures. The nonmonotonous behavior
close to the Kondo temperature is dominated by correlation effects, while the
interference effects lead to the monotonous increase of the differential
conductance with $b_1$ at higher temperatures. 
 \label{figdIdV_b1T}}
\end{figure}

Further calculation is performed for symmetric couplings to the source and to
the drain, that is for $t_{i k}^s=t_{i k}^d$. In that case the tunneling
angles of each mode in the source and in the drain leads are equal, $\eta_{
k}^s=\eta_{k}^d$. Furthermore, as it has already been done in calculations
of the Kondo temperature (see Section \ref{secT}), we neglect a possible
dependence of the total tunneling to the double quantum dot on the absolute
value of the wave vector $\vert k\vert$. Then, as follows from the definition
(\ref{PsiXi}), the
components of the Green's function $\mathcal{F}^{R/A}$ vanish, and the only
nonzero component of the function $\mathcal{F}$ reads 
\begin{eqnarray}
\nonumber && 
\mathcal{F}^K(\epsilon, k) =(-2\pi i)\frac{t^s
t^d}{t^2} \left\{\tanh\left(\frac{\epsilon-\mu_s}{2T}\right) 
\right. \\ 
&& \left. 
-\tanh\left(\frac{\epsilon-\mu_d}{2T}\right)\right\}
\delta(\epsilon-\xi_{k}). 
\label{F^K}
\end{eqnarray}
Performing the evaluation of the diagram in Fig. \ref{current}, we obtain the   
expression for total current through the double quantum dot in the form 
\begin{equation}
j(V)=40\pi e^2 V \nu_0^2 \frac{\left(t^s t^d\right)^2}{t^4}
\sum_{\lambda,\mu,\nu} J_{\lambda\mu\nu}^2 \left(\frac{1}{2} +2b_1^2\right). 
\label{j_final} 
\end{equation} 
Here $\nu_0$ denoted the one particle density of states in the source and drain leads. 
Let us comment on the origin of different factors in (\ref{j_final}). The
factor $\sum_{\lambda \mu \nu} J_{\lambda\mu\nu}^2$, where the sum runs over the
whole set of the Kondo coupling constants generated by the RG transformation,
reflects the influence of the spin and orbital Kondo correlations on the current
through the double dot.  In contrast, the term $\frac{1}{2}
+2b_1^2$ describes the single particle interference effects by transport
through the double dot device. Namely, in the case $b_1=0$, when each dot has
its own reservoir of modes in the source and in the drain lead, the current
through the device is obtained as a sum of the currents through the dot 1 and
through the dot 2 (see Fig. \ref{fig-dd-channels}a). In the other limit case
$b_1=1/2$, there is a common
reservoir for both dots. Then a typical geometry of a double-slit experiment is
realized. Due to the interference effect, the current in the interference
maximum is doubled with respect to its classical value.  Indeed, the value of
the factor $\frac{1}{2} +2b_1^2$ is twice as much for $b_1=1/2$
as for $b_1=0$ (see Fig. \ref{fig-dd-channels}b).  At the intermediate values of
$b_1$ the single particle coherence is partially suppressed.  
The correlation
factor dominates the
dependence of the differential conductance at temperatures close to the Kondo
temperature, whereas the single particle interference determines the behavior of
conductance at higher temperatures further away from $T_K$. These findings are
illustrated in Fig. \ref{figdIdV_b1T}. According to the behavior of the
conductance as a function of the mixing parameter $b_1$, the low-temperature and
the high-temperature regimes can be identified that are dominated by the many
particle corelations or by the single particle interference respectively.  

\begin{figure}
 \hskip -5 cm a)\\
\hskip -1cm \includegraphics[width=7cm,height=6cm,angle=0]{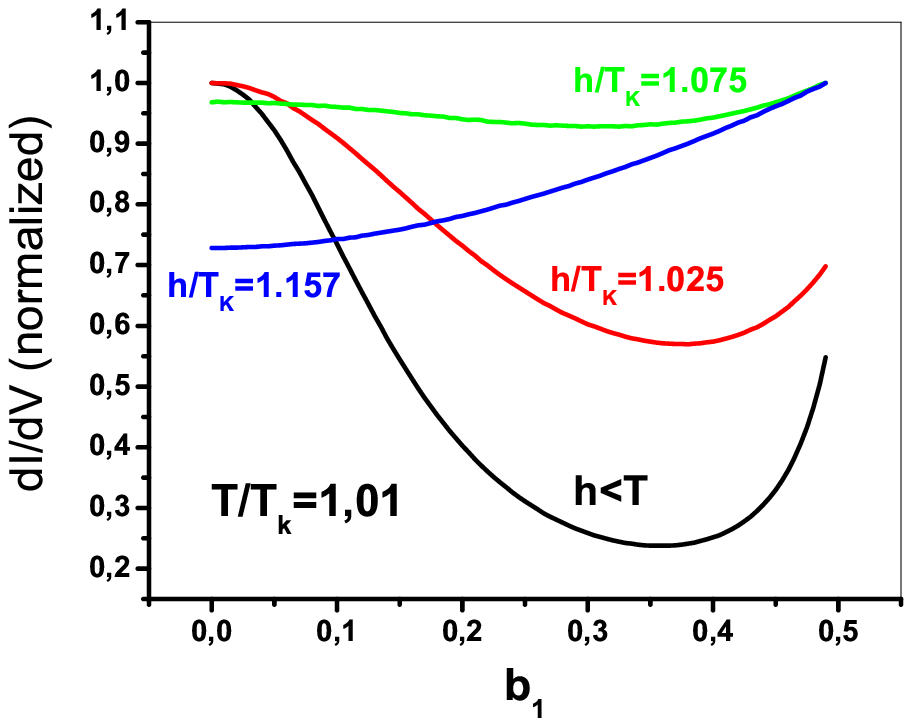}\\ 
\hskip -5 cm b)\\
\hskip -1cm\includegraphics[width=7cm,height=6cm,angle=0]{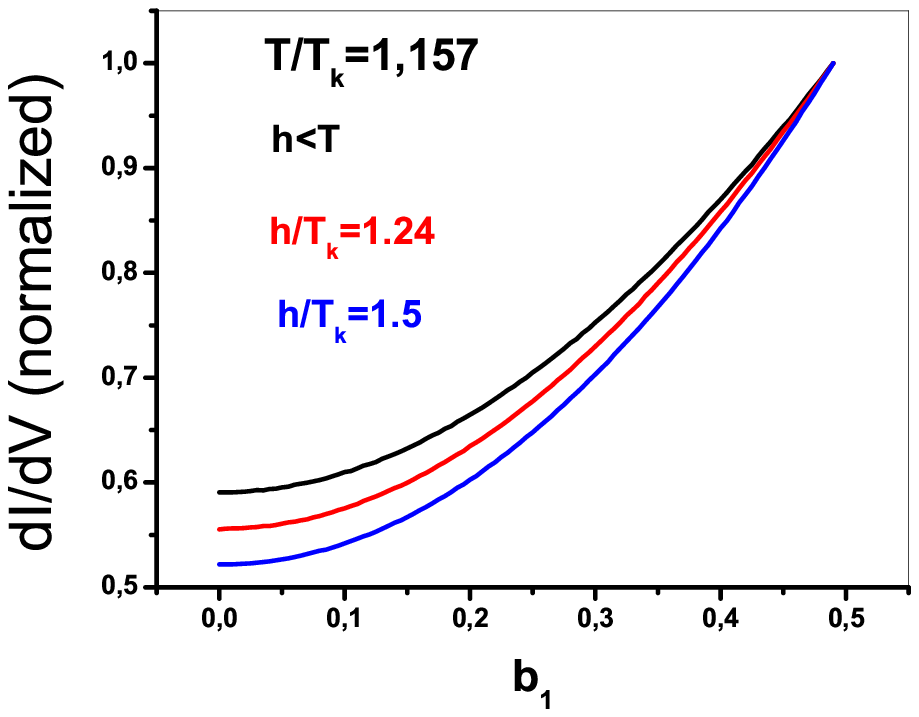}
\caption{(Color online) Differential conductance of the double quantum dot 
as a function of the mixing parameter $b_1$ at
different Zeeman fields. a) Low temperature regime:  The change from
nonmonotonous behavior to the monotonous increase is due to  the suppression of
Kondo correlations by Zeemann field. b) High temperature regime: The behavior
of conductance is dominated by the one-particle interference.  
 \label{figdIdV_zeeman}}
\end{figure}

Suppressing the Kondo correlations in the spin sector by a Zeeman magnetic 
field, one extends the regime, where the single particle interference dominates
the behavior of conductance. At finite Zeeman magnetic field, the regimes of
high and low field can be identified according to the qualitative behavior of
the conductance, as shown in Fig. \ref{figdIdV_zeeman}.  In the high
temperature regime, the conductance increases monotonously with $b_1$ at all
Zeeman fields. Increasing the Zeeman field  affects only the
conductance at small values of $b_1$ suppressing it. At $b_1$ close to 0.5 and
Zeemann field, both spin and pseudospin Kondo correlations are suppressed, and
the conductance is determined by the single particle interference factor only. 

\section{Conclusion \label{sec-conclusions}}

In this work we considered the transition between the SU(4) and SU(2) Kondo
effects in a double dot system embedded in a parallel circuit with attached
leads. The transition is induced by the transfer of electrons between that dots
through the leads, which violates the conservation of the pseudospin. Despite
having the same Kondo temperature, the SU(4) and spin SU(2) Kondo effects can be
distinguished by the sensitivity to the in plane Zeeman magnetic field. We
derived the dependence of the Kondo temperature on Zeeman magnetic field
(\ref{T(h)})
that allows direct comparison with experiments. Being the basic energy scale,
the Kondo temperature determines all other experimentally observable properties
related to the Kondo effect. In particular, the behavior of the differential
conductance through the double quantum dot as a function of the mixing parameter
$b_1$
is determined by the interplay of two factors, the suppression of the
pseudospin 
 Kondo correlations with $b_1$ and the increase of the single particle
coherence. Domination of the correlation factor leads to the dependence of
conductance that is similar to the dependence of the Kondo temperature on
$b_1$, while the dominance of the interference factor leads to the monotonous
increase of the differential conductance with the mixing parameter $b_1$. 

One can propose several ways to change the mixing between the reservoirs in
experiment. Starting from the case of fully mixed reservoirs $b_1=0.5$ the
separation can be achieved by diminishing  the single particle coherence length.
As soon as the coherence length is smaller than the separation between the two
dots, the coherent propagation of electrons between the two dot through the
source or through   the drain leads becomes suppressed, and each dot is coupled
to its own subset of modes, which corresponds to the case $b_1=0$ 
(see Fig. \ref{fig-dd-channels}a). 

Therefore, the calculated dependence of the Kondo temperature  and of the
differential conductance on Zeeman magnetic field and on the mixing parameter
can be seen in the measurements of the conductance through the double quantum
dot device. The obtained results can also be relevant to the experiments on
Kondo effect in carbon nanotubes \cite{Choi}.

\begin{acknowledgments}
A.C. and F.H.  thank D. Pfannkuche and M. Trushin for valuable comments.
Support from Deutsche Forschungsgemeinschaft through Sonderforschungsbereiche
SFB 508 (F. H., V. K.)  and SFB 668 (A. C.) is gratefully acknowledged. 
\end{acknowledgments}

\end{document}